\documentclass[11pt]{article}
\usepackage{moriond,epsfig}

\def\Journal#1#2#3#4{{#1} {\bf #2}, #3 (#4)}

\def\NPB{{\em Nucl. Phys.} B}
\def\PLB{{\em Phys. Lett.}  B}

\def\PRD{{\em Phys. Rev.} D}

\def\be{\begin{equation}}
\def\ee{\end{equation}}
\def\bea{\begin{eqnarray}}
\def\eea{\end{eqnarray}}

\begin{document}
\vspace*{4cm}
\title{O.P.E. AND POWER CORRECTIONS ON THE Q.C.D. COUPLING CONSTANT}

\author{PH. BOUCAUD, F. DE SOTO, A. DONINI, J.P. LEROY, A. LE YAOUANC, J. MICHELI, \\ H. MOUTARDE, O. P\`ENE, 
\underline{J. RODR\'IGUEZ-QUINTERO}}

\address{L.P.T., Universit\'e de Paris XI, B\^atiment 210, 91405 Orsay Cedex, France \\
Dpto. de F.A., E.P.S. La R\'abida, Universidad de Huelva, 21819 Palos de la fra., Spain }

\maketitle\abstracts{The running coupling constant can be estimated by computing 
gluon two- and three-point Green functions from the lattice. Computing in lattice implies 
working in a fixed gauge sector (Landau). An source of systematic uncertainty is then
the contribution of non gauge-invariant condensates, as $\langle A^2 \rangle$, generating
power corrections that might still be important at very high energies. We study the impact of 
this gluon condensate on the analysis of gluon propagator and vertex lattice data and on 
the estimate of $\Lambda_{\overline{\rm MS}}$. We finally try a qualitative description of
this gluon condensate through the instanton picture.}

\section{The running coupling constant from the lattice}

That the running coupling constant can be extracted from the three-gluon vertex in the 
Landau gauge was proposed several years ago in a seminal work~\cite{Par}. The key lies on 
the appropriate choice of renormalisation scheme: the so-called Momentum Substraction (MOM) schemes 
are defined such that all the renormalised Green functions take their tree-level expressions
after replacing bare by renormalised constants. Then

\bea
g_R(k^2)  & = & \frac{G^{(3)}(k_1^2,k_2^2,k_3^2)}{\left(G^{(2)}(k^2)\right)^3} \
\left(Z^{\rm MOM}(k^2)\right)^{3/2} 
\eea

\noindent where $G^{(2)}$ and $G^{(3)}$ are the scalar form factor of two- and three-gluon 
Green functions and $Z^{\rm MOM}$ is the gluon propagator renormalisation constant. Alternative 
kinematics for the renormalisation point are possible, mainly two among them: symmetric 
($k_1^2=k_2^2=k_3^2=k^2$) and asymmetric ($k_1^2=k_2^2=k^2$,$k_3^2=0$).
Analysis of lattice computations of these Green functions renormalised in both yielded~\cite{alpha} the estimates
of the coupling and $\Lambda_{\overline{\rm MS}}$ collected in Table \ref{tab1}. 
Concerned as we were by the determination and renormalisation of the gluon propagator for the 
analysis of the three-point Green Function, we exploited~\cite{propag} the first by matching the data to the three-loop perturbative 
prediction and hence estimate a purely perturbative coupling and $\Lambda_{\overline{\rm MS}}$. The disagreement 
between estimates from both gluon vertex and propagator (see Table \ref{tab1}) manifests the impact of some ucontrolled systematic 
uncertainty. Even worse, the so estimated $\Lambda_{\overline{\rm MS}}$ does not behave as a scale invariant!!

\begin{table}[hbt]
\caption{ Estimates of the coupling and $\Lambda_{\overline{\rm MS}}$ 
from two- and three-point Green function methods .\label{tab1}}
\vspace{0.4cm}
\begin{center}
\begin{tabular}{|c|c|c|}
\hline
& & \\
&
Three-point &
Two-point
\\ \cline{2-3}
$\alpha(4.3 \rm{GeV})$ & 0.269(3) &  \\
\hline
$\Lambda_{\overline{\rm MS}}(4.3 \rm{GeV})$ & 299(7) & \\
\hline
$\alpha(9.6 \rm{GeV})$ & 0.176(2) &  0.193(3) \\
\hline
$\Lambda_{\overline{\rm MS}}(9.6 \rm{GeV})$ & 266(7) & 319(14) \\
\hline
\end{tabular}
\end{center}
\end{table}

\begin{figure}[hbt]
\begin{center}
\rule{5cm}{0.2mm}\hfill\rule{5cm}{0.2mm}
\vskip 0.5cm
\begin{tabular}{cc}
\epsfig{figure=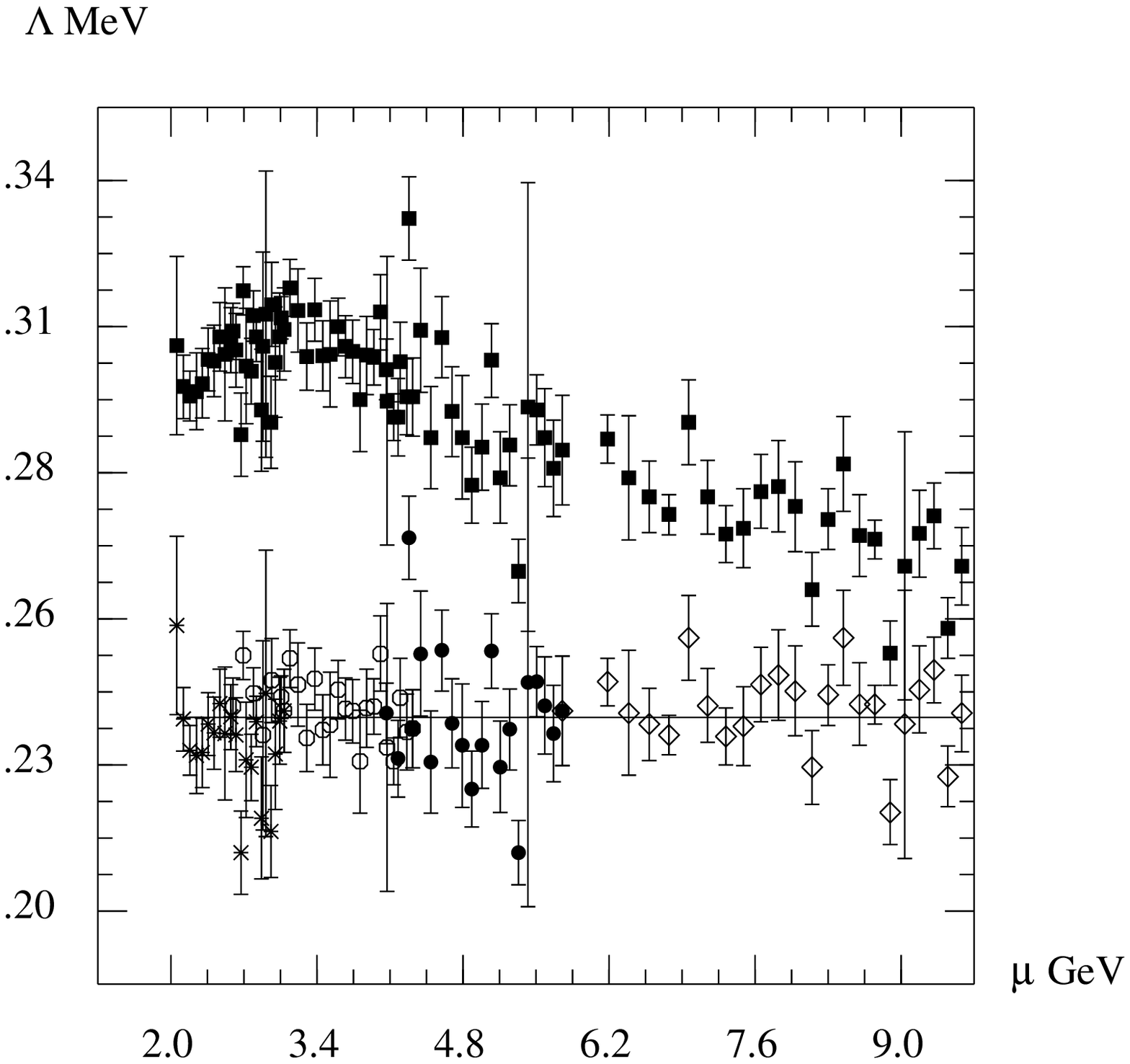,height=2.5in} &
\epsfig{figure=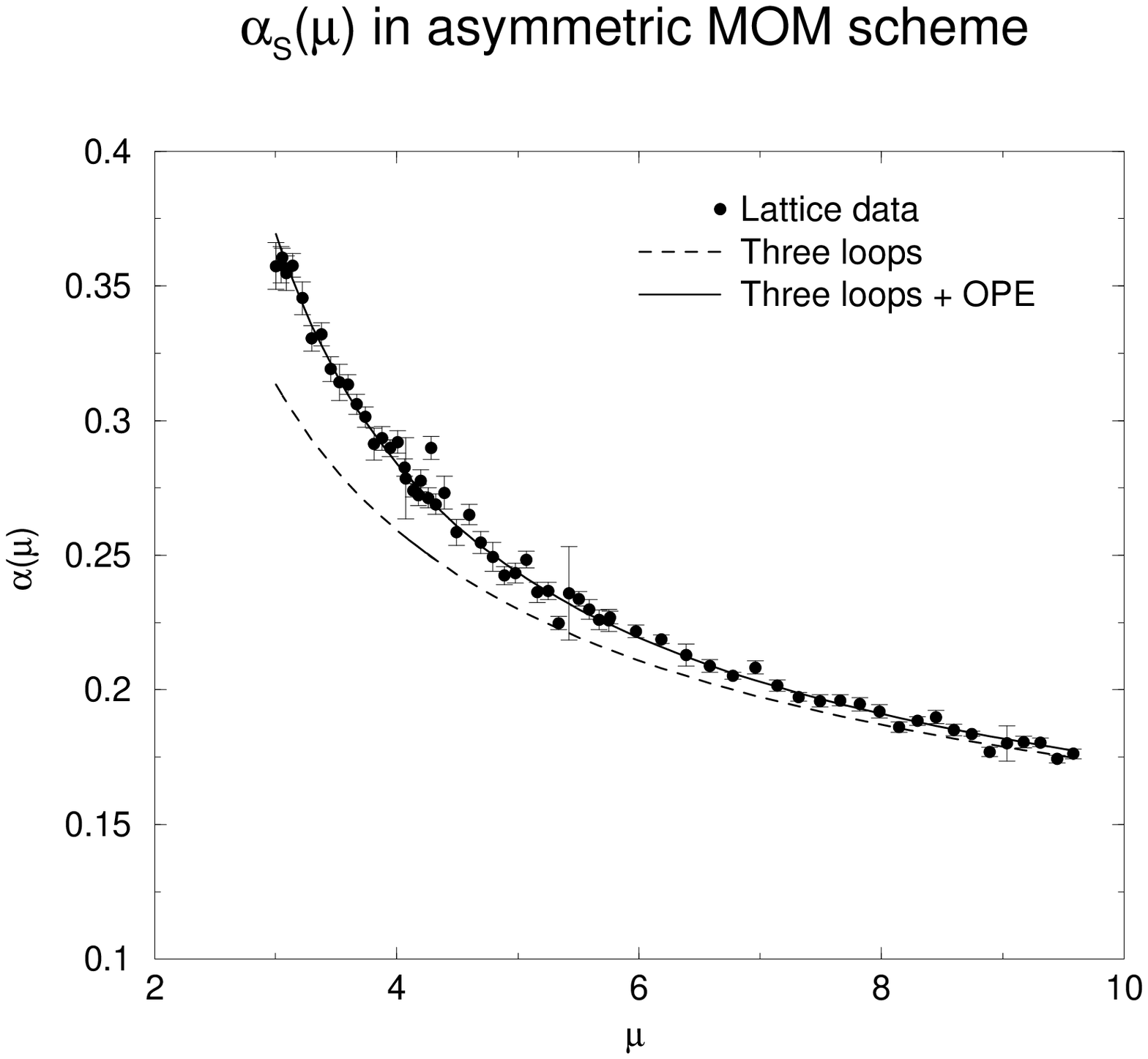,height=2.5in} \\
(a) & (b) \\
\end{tabular} \\
\rule{5cm}{0.2mm}\hfill\rule{5cm}{0.2mm}
\end{center}
\caption{In plot (a), the upper solid points are for the estimates of $\Lambda_{\overline{\rm MS}}$
obtained through the non perturbative couplings directly computed from lattice Green functions. 
The lower ones are obtained through $\alpha_s^{\rm pert}$ in Eq. \ref{eq:G2alpha2} by the substraction
of the fitted power contribution. Plot (b) shows the best fit of lattice data for the coupling renormalised 
in asymmetric MOM to Eqs. \ref{eq:G2alpha2}, \ref{eq:Cs}. 
\label{fig1}}
\end{figure}

If one empirically adds a $1/p^2$ power correction to the perturbative formulae (we work at three loops)

\bea
G(p^2)= G(p^2)_{\rm 3 loops}\left(1 + \frac c {p^2}\right) \  \  \  , \ \ \  
\alpha_s(p^2)=\alpha_s(p^2)_{\rm 3 loops}\left(1 + \frac {c'} {p^2}\right) \ ,
\label{eq:G2alpha2}
\eea

\noindent success is gained both in obtaining estimates by two- and three-point methods that agree to each other and 
in restoring the scale invariance of the estimated $\Lambda_{\overline{\rm MS}}$ parameter~\cite{poweral} (see 
fig. \ref{fig1}.a). A third outcome arises from matching to Eq. \ref{eq:G2alpha2}: {\it the estimate of 
$\Lambda_{\overline{\rm MS}}$ results}  {\bf 237(4) MeV } (only the statistical error is quoted), {\it in astonishing 
agreement with Schr\"oedinger functional's~\cite{Sommer}:} {\bf 238(19) MeV } !!
 
\section{The O.P.E. picture}

\begin{figure}
\begin{center}
\rule{5cm}{0.2mm}\hfill\rule{5cm}{0.2mm}
\vskip 0.5cm
\epsfig{figure=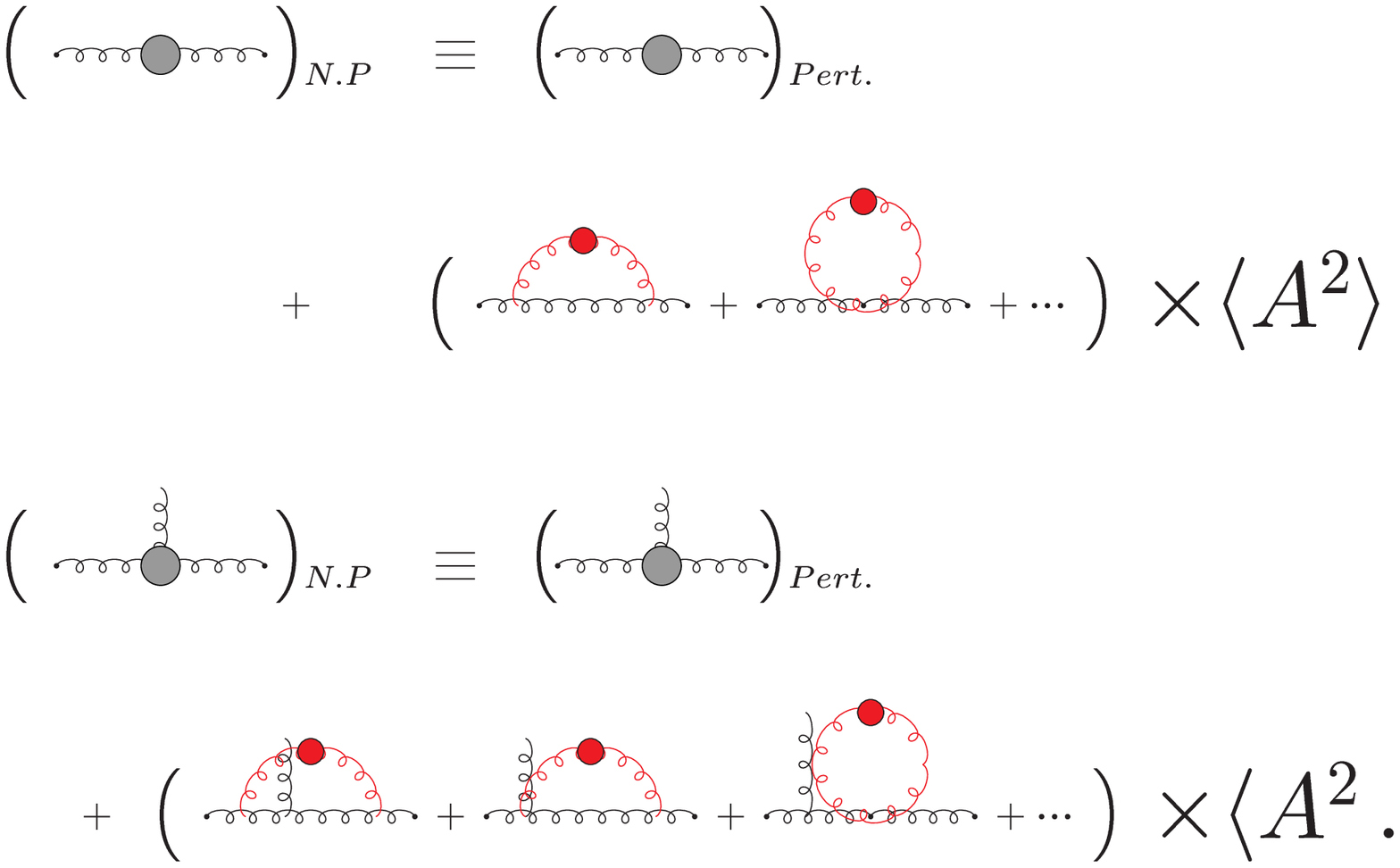,height=2.3in} 
\\
\rule{5cm}{0.2mm}\hfill\rule{5cm}{0.2mm}
\end{center}
\caption{ Diagrams in the OPE expansion of gluon two- and three-point Green functions.
\label{fig2}}
\end{figure}

How large should $c$ and $c'$ are? The Operator Product Expansion (0PE) can be invoked
to let us gain some physical insight into the matter of these coefficients. The sum rule approach connects
the power corrections to any non-local matrix element and the QCD vacuum expectation 
value of the local operators from the expansion, the so-called {\bf condensates}, and gives the prescription 
to compute perturbatively the coefficients of the expansion (Wilson coefficients). Symmetries and power 
counting tell us that the gluon condensate $\langle A^2 \rangle$ gives the major contribution to our gluon 
Green functions. This is of course a non gauge-invariant contribution but, as far as we work in a fixed gauge, 
Landau gauge, non-vanishing contributions from gauge-dependant condensates, {\it i.e.} $\langle A^2 \rangle$, 
should be expected~\cite{lavelle}. Then, our coefficients $c$ and $c'$ can be computed~\cite{opes} (see fig. \ref{fig2} )
and written in terms of the gluon condensate renormalised at any momentum scale $\mu$,

\bea
c'=\chi \ c \ \ ,\ \ \ c \ = \frac{6 \pi^2}{\beta_0 (N_c^2-1)} \left(
\ln{\frac{p}{\Lambda}} \right)^{\frac{\gamma_0+\widehat{\gamma}_0}{\beta_0}-1} \
\langle A^2 \rangle_{\mu} \left(
\ln{\frac{\mu}{\Lambda}} \right)^{-\frac{\gamma_0+\widehat{\gamma}_0}{\beta_0}} \ ;
\label{eq:Cs}
\eea

\noindent with $\gamma_0=13/2, \beta_0=11, \widehat{\gamma}_0= 3 N_C/4$. The factor $\chi$ depends on the 
particular kinematics we choose for the vertex renormalisation point; case asymmetric: $\chi=1$, case 
symmetric: $\chi=3$. 

Then, Eqs. \ref{eq:G2alpha2}, \ref{eq:Cs} are at our disposal for trying a coherent description of 
gluon propagator and vertex lattice data, with the sole necessity of fitting two ingredients of 
non-perturbative nature: $\langle A^2 \rangle$ and $\Lambda_{\overline{\rm MS}}$ (except for 
an overall factor in the propagator analysis). We expand purely perturbative series up to the third loop and only keep 
the leading logarithm contribution for the Wilson coefficient. Of course this does not guarantee us to reach the 
asymptotic regime where expansions well behave. Thus, we approach the problem through a combined analysis of both gluon
propagator and vertex, where we look for fitting the same $\Lambda_{\overline{\rm MS}}$ parameter in both and two independent 
estimates of $\langle A^2 \rangle$ to be compared. The results of applying this to our lattice data~\cite{opes} 
are in Table \ref{tab2}. The best fit parameters lead, for instance in the case of the MOM asymmetric vertex, to the 
plot (b) in fig. \ref{fig1}.

\begin{table}[hbt]
\caption{\small Comparison of results obtained to the renormalisation momentum scale $\mu=10$ GeV.}
\vspace{0.4cm}
\begin{center}
\begin{tabular}{|c|c|c|}
\hline
 & asymmetric MOM & symmetric MOM \\
 \hline
$\Lambda_{\overline{\rm MS}}$ & 260(18) MeV & 233(28) MeV\\
\hline
$\left\{\sqrt{\langle A^2 \rangle_{R,\mu}}\right\}_{prop} $ &  1.39(14) GeV & 1.55(17) GeV\\
\hline
$\left\{\sqrt{\langle A^2 \rangle_{R,\mu}}\right\}_{alpha}$ & 2.3(6) GeV & 1.9(3)GeV \\
\hline
\end{tabular}
\label{tab2}
\end{center}
\end{table}

Consequently, the contribution of this non gauge-invariant condensate $\langle A^2 \rangle$ seems to explain 
rather well the systematic deviation in the matchings of our gluon propagator and vertex lattice data 
to purely perturbative formulae.

\section{Instantons and $\langle A^2 \rangle$ condensate}

The physical origin of this gluon condensate is a major question. A common belief is that an 
instanton ensemble (liquid or gas) provides with a fair description of important features of the QCD
vacuum. Then, if one considers a hard gluon of momentum $p_\mu$ propagating in an instanton gas 
background (we crudely assume that instantons do not interact to each other), the propagator can 
be computed with Feynman graphs and it is easy to see that the dominant contribution coming from its 
interaction wiht the instanton gauge field is $O(1/p^2)$ when the gauge field momentum, 
$k_\mu << p_\mu$. This correction is equal 
to the standard OPE Wilson coefficient for the propagator times

\be
\langle A^2_{\rm inst} \rangle = \frac{n_I+n_A}{V} \int d^4x \sum_{\mu,a} A_\mu^a A_\mu^a = 
12 \pi^2 \rho^2 \frac{n_I+n_A}{V} 
\label{eq:A2I}
\ee

\noindent where $A_\mu^a$ is the standard `t Hooft-Polyakov solution in the singular Landau gauge, $n_I(n_A)$ 
is the number of (anti-)instantons and $\rho$ is the instanton radius. We estimate~\cite{instanton} this instanton-induced 
condensate to be 1.76(23) GeV$^2$ by performing several simulations at $\beta=6.0$ on 
a $24^4$ lattice, applying the cooling procedure to them for killing UV fluctuations and computing the 
instanton density to be applied in Eq. \ref{eq:A2I}. This result is 
to be compared with the OPE estimate from the propagator analysis. We should first
run down~\footnote{In the instanton analysis, the renormalisation is performed through the cooling
that kills the UV fluctuations. The only new physical momentum scale related to the instanton background emerging after
the cooling procedure is the inverse of the instanton radius $\sim 0.5$ GeV} the value in Table \ref{tab2} to the lowest renormalisation momentum 
scale we still believe for our OPE analysis, 
that is $\sim 2.5$ GeV. We obtain 1.4(3) GeV$^2$. This fair agreement seems to indicate that the 
gluon condensate receives a significant 
instantonic contribution.

\end{document}